\documentclass[prl,showpacs,twocolumn]{revtex4}
\usepackage{amsmath}
\usepackage{graphicx}
\begin{document}
\title{Robustness of topologically protected surface states in layering of Bi$_2$Te$_3$
thin films}
\author{Kyungwha Park, J. J. Heremans, V. W. Scarola, and Djordje Minic}
\affiliation{Department of Physics, Virginia Tech, Blacksburg, Virginia 24061}
\begin{abstract}
Bulk Bi$_2$Te$_3$ is known to be a topological insulator. We investigate surface states of 
Bi$_2$Te$_3$(111) thin films using density-functional theory including spin-orbit coupling. 
We construct a method to unambiguously identify surface states of thin film
topological insulators. Applying this method for one to six quintuple layers of Bi$_2$Te$_3$,
we find that the topological nature of the surface states remains robust with the film
thickness and that the films of three or more quintuple layers have topologically non-trivial
or protected surface states, in agreement with recent experiments.
\end{abstract}
\date{\today}

\maketitle


Recently, topological insulators (TIs) with time-reversal symmetry have attracted attention due to
their topologically protected states \cite{FU07,TEO08,QI08,HASA10}.
In three dimensions, TIs differ
from band insulators in that a bulk energy gap opens due to strong spin-orbit coupling (SOC)
with metallic surface states in the bulk energy gap.
Several bulk bismuth-based alloys were discovered to be three-dimensional
TIs \cite{TEO08,QI08,HASA10,ZHANG09,HSIE08,CHEN09,HSIE09-3,XIA09,HSIE09-2,ZHANG09-T,ROUS09,GOME09}.
For ordinary metals, surface states are so fragile that they can be destroyed by a small number of impurities or
disorder at the surface. However, the surface states of TIs are topologically protected in that
impurities preserving time-reversal symmetry can neither destroy nor impact the topological
nature of the surface states.

Three-dimensional TIs are classified according to a topological invariant, the Z$_2$ 
invariant $\nu_0$ \cite{FU07}. 
Strong (weak) TIs have $\nu_0=1$ ($\nu_0=0$). 
Recent first-principles calculations \cite{ZHANG09,ZHANG10-W} show that bulk Bi$_2$Te$_3$, Bi$_2$Se$_3$, and
Sb$_2$Te$_3$ alloys are strong TIs with a single Dirac cone below the Fermi level, $E_f$, at $\Gamma$ ($\vec{k}=0$).
This feature was confirmed by angle-resolved photoemission spectra (ARPES)
experiments \cite{HASA10,CHEN09,HSIE09-3,XIA09,HSIE09-2,ZHANG09-T}. 

Thin films offer valuable probes of TIs as well as potential device applications. It is, therefore, important to 
identify surface states and their topological properties. For example,
thin TI films were proposed to be efficient thermoelectric devices that exploit the interaction between top and
bottom surface states \cite{GHAE09}. Thin films also have considerable advantages in a direct measurement of transport 
properties of the surface states by emphasizing surface states over bulk states. Very recently,
Bi$_2$Te$_3$ and Bi$_2$Se$_3$ thin films with a thickness of a few nm were experimentally realized
\cite{ZHANG09-G,TEWE10,ZHANG09-Y,LI09,SAKA10}, and topological properties of such films have
been examined \cite{ZHANG09-Y,LI09,SAKA10,LIU10,LIND09,LU10}. One theory suggests that the quantum spin 
Hall phase of a thin TI film oscillates between topologically trivial ($\nu_0=0$) and nontrivial ($\nu_0=1$)
states with the film thickness \cite{LIU10}, while another study implies a topological quantum phase 
transition with an oscillation in an energy gap $\Delta$ with the thickness \cite{LU10}. However, ARPES measurements 
\cite{ZHANG09-Y,LI09,SAKA10} on thin films did not show any oscillation in either $\nu_0$ or $\Delta$
with thickness. These discrepancies cast doubt on the robustness of topological surface states and how to
identify them.

In this work, we construct a method to unambiguously identify topologically protected surface states for thin TI 
films. Our method based on density-functional theory (DFT) with SOC, can be used to determine the 
topological nature of the surface states in thin TI films. As an example, we apply this method to Bi$_2$Te$_3$
thin films at six different thicknesses. We show that the topological nature of the surface states remains 
{\it robust} with the film thickness and that the surface states are topologically non-trivial ($\nu_0=1$) for 
films three or more quintuple layers (QLs) thick. This differs from previous calculations \cite{LIU10}, yet
agrees with experiment \cite{LI09}. 

\begin{figure}
\includegraphics[width=4.1cm, height=3.2cm]{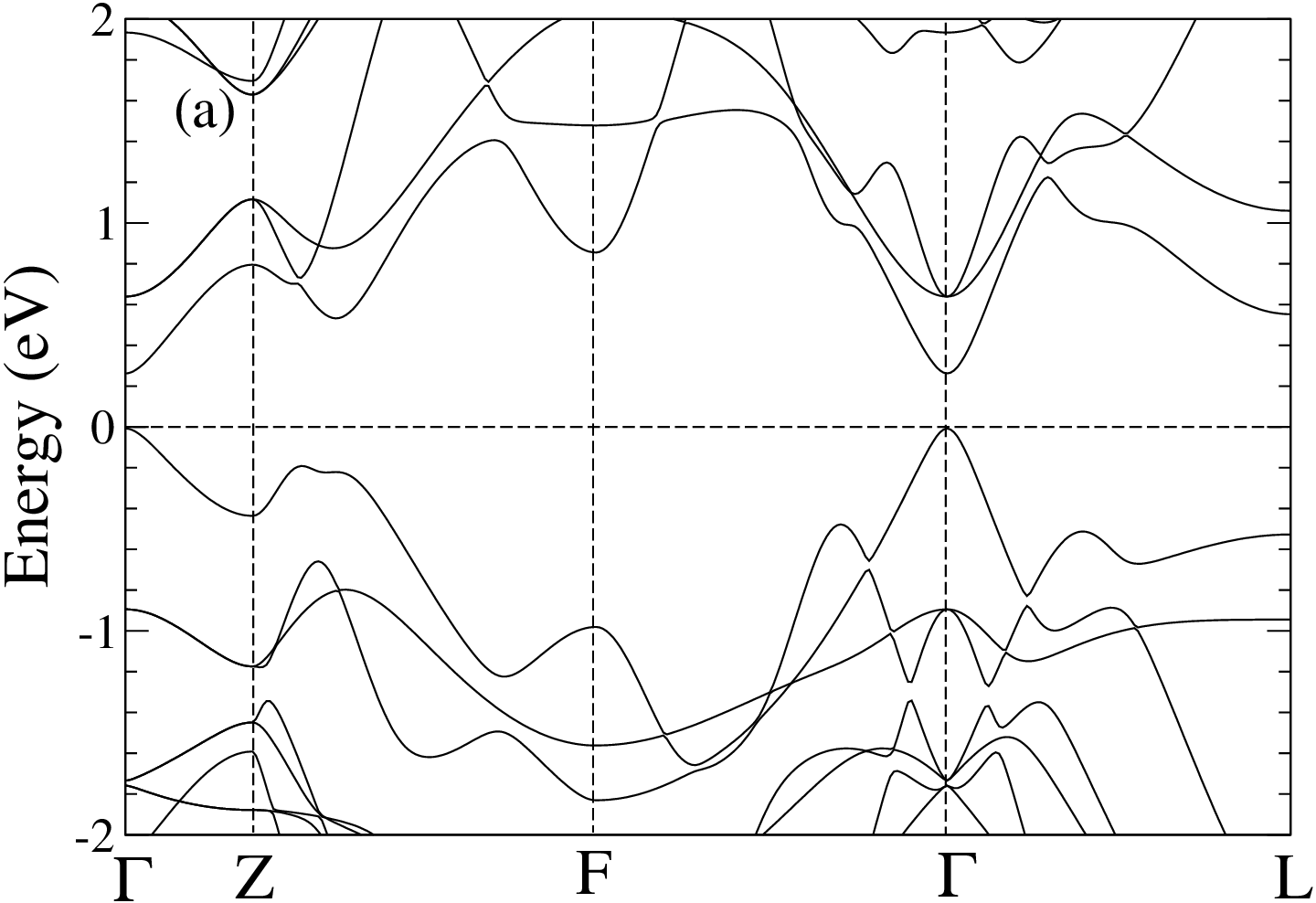}
\hspace{0.1truecm}
\includegraphics[width=4.1cm, height=3.2cm]{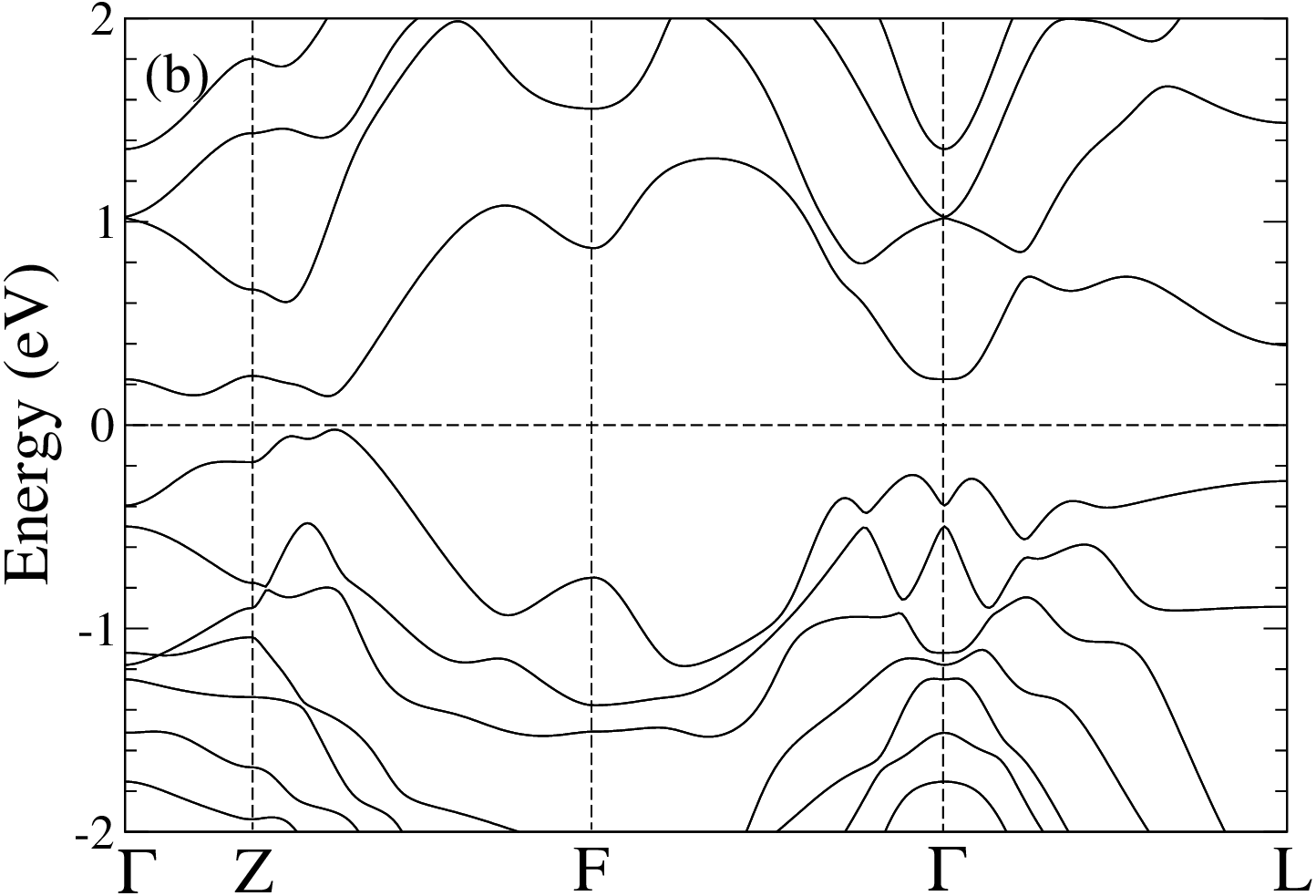}
\caption{Band structure of bulk Bi$_2$Te$_3$ (a) without SOC and (b) with SOC at
symmetry points $\Gamma$, Z, F, and L.}
\label{fig:bulk-band}
\end{figure}

We begin with a review of the bulk properties of Bi$_2$Te$_3$ \cite{MISH97}. A unit cell of bulk Bi$_2$Te$_3$ 
consists of a rhombohedral structure with five inequivalent atoms: two Bi and three Te. For the lattice
constants ($a$=4.386~\AA~and $c$=30.497~\AA)~and the positions of the five atoms, we use experimental 
data \cite{NAKA63}. We calculate the electronic structure of
bulk Bi$_2$Te$_3$ using a DFT code, {\tt VASP} \cite{VASP}, within the Perdew-Burke-Ernzerhopf (PBE)
generalized-gradient approximation (GGA) \cite{PERD96}. Projector-augmented-wave (PAW) pseudopotentials are
used \cite{PAW}. Plane waves with a kinetic energy cutoff $E_c$ of 175~eV are used as basis sets
and 146 irreducible $k$-points ($N_k$=146) are sampled.
Figure~\ref{fig:bulk-band} shows the bulk band structure computed without and with SOC at time-reversal
invariant momenta, $\Gamma$, $Z$, $F$, and $L$, with $E_f=0$. At the Dirac point ($\Gamma$), SOC inverts
the conduction and valence bands with opposite parities \cite{ZHANG09}.

\begin{table}
\begin{center}
\caption{The number of crossings $N_c$ of the surface states across $E_f$, the energy gap
$\Delta$ at $\overline{\Gamma}$, and the indirect energy gap $\Delta E_{\rm ind}$ (eV)
as a function of slab thickness (nm). $E_c$ (eV) and $N_k$ are the energy cutoff and
the number of $k$ points sampled, respectively.}
\label{table:1}
\begin{ruledtabular}
\begin{tabular}{c|c|c|c|c|c|c}
no of QLs & thickness      & $E_c$      & $N_k$ & $\Delta$ (eV) &  $N_c$  & $\Delta E_{\rm ind}$ \\ \hline
   1      &  1.0166        &  300       & 48    &  0.4338       &  0               & 0.301 \\
   2      &  2.0332        &  300       & 96    &  0.1319       &  0               & 0.057 \\
   3      &  3.0497        &  500       & 341   &  0.0261       &  1               & No gap   \\
   4      &  4.0663        &  500       & 341   &  0.0070       &  1               & No gap   \\
   5      &  5.0829        &  500       & 341   &  0.0090       &  1               & No gap   \\
   6      &  6.0993        &  500       & 341   &  0.0055       &  1               & No gap
\end{tabular}
\end{ruledtabular}
\end{center}
\end{table}

\begin{figure*}
\includegraphics[width=5.6cm, height=4.4cm]{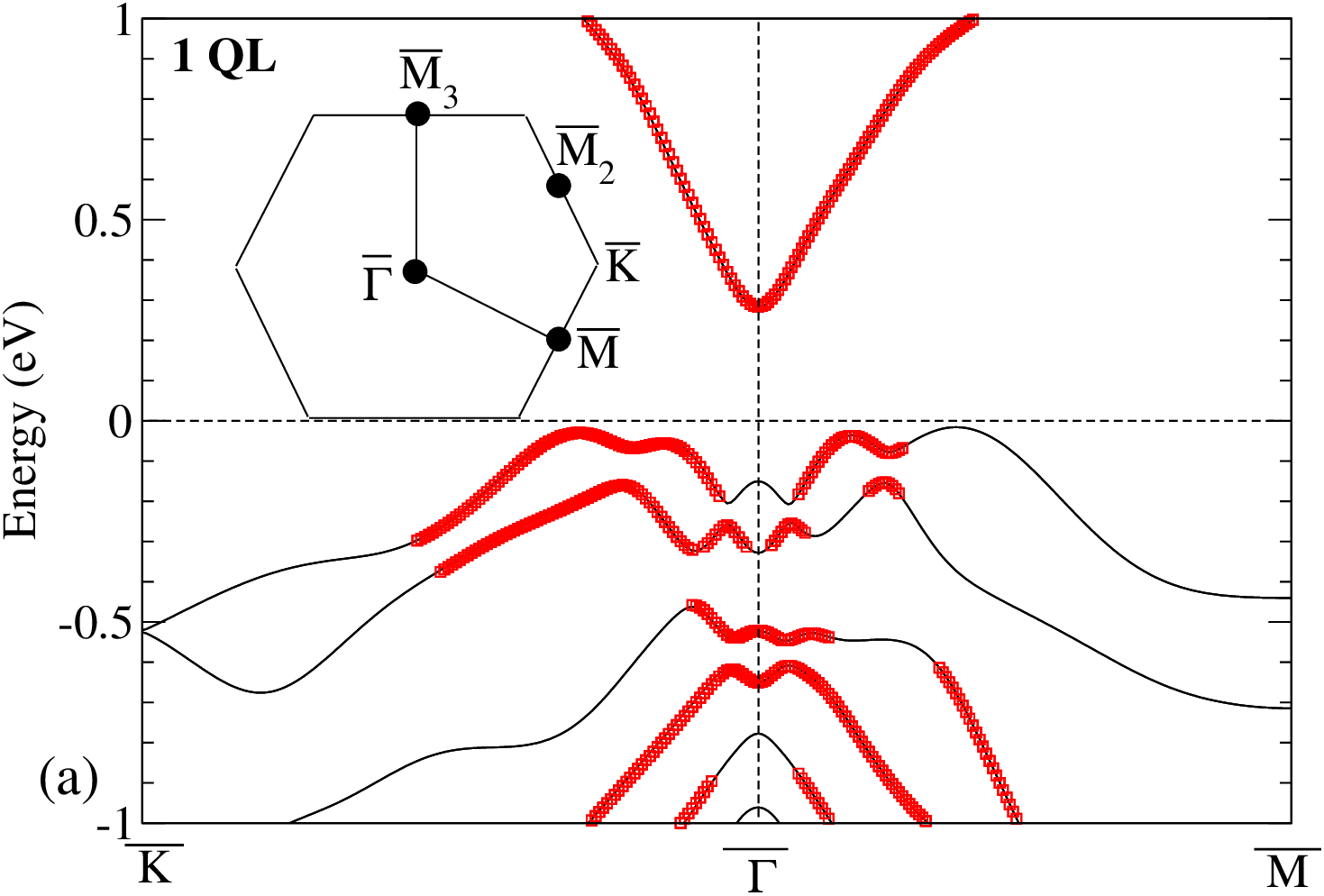}
\hspace{0.2truecm}
\vspace*{0.4truecm}
\includegraphics[width=5.6cm, height=4.4cm]{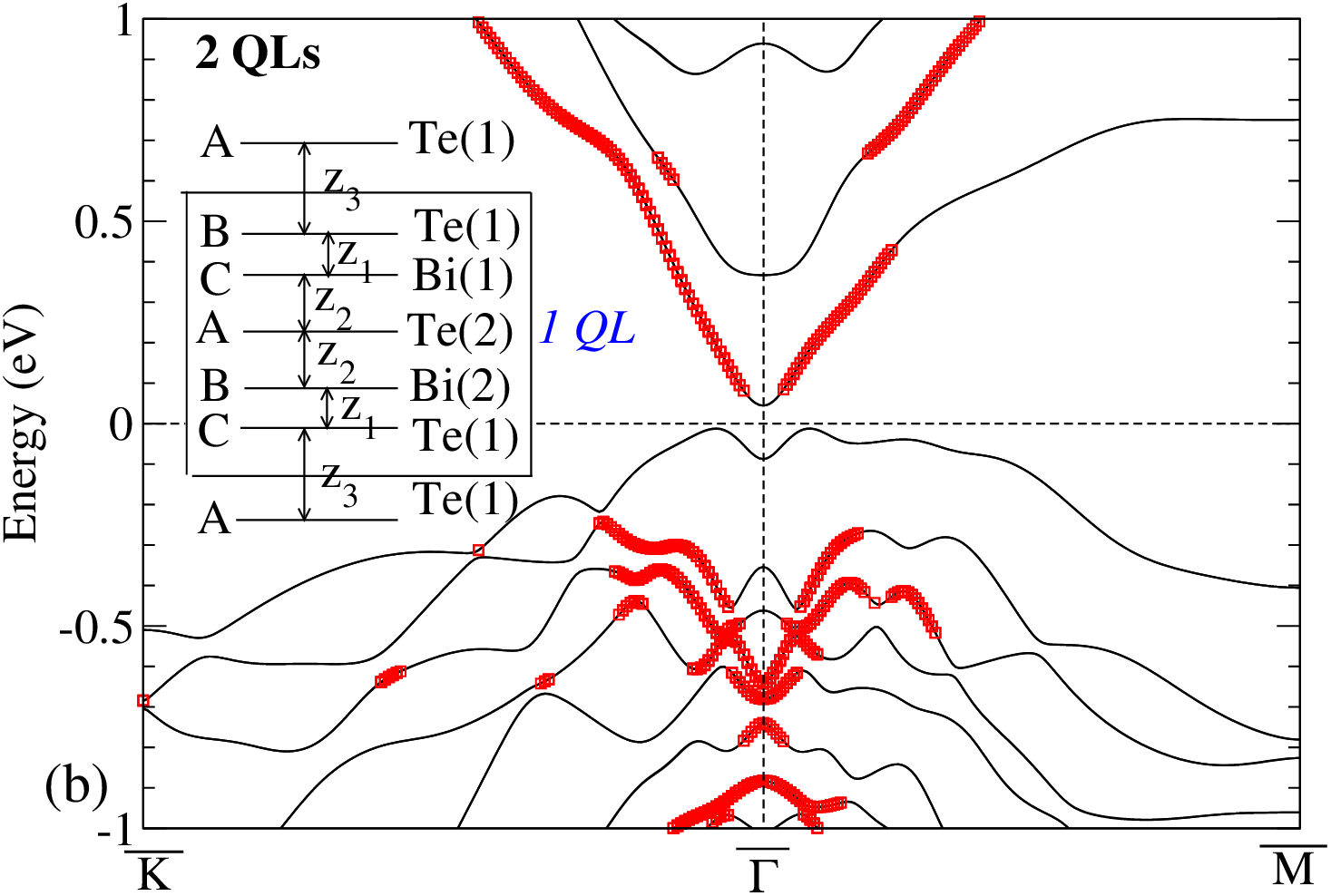}
\hspace{0.2truecm}
\vspace*{0.4truecm}
\includegraphics[width=5.6cm, height=4.4cm]{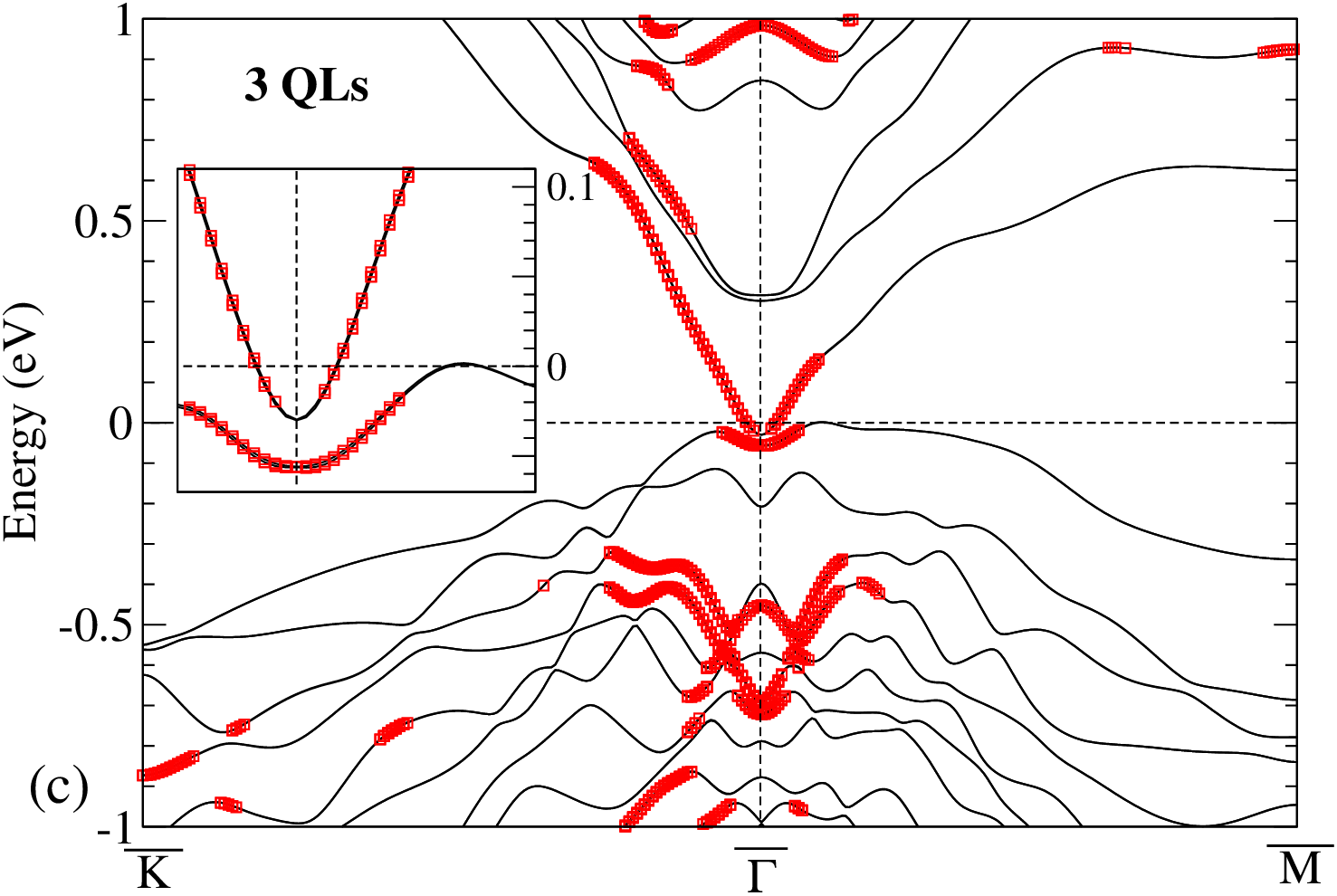}
\vspace*{0.4truecm}
\includegraphics[width=5.6cm, height=4.4cm]{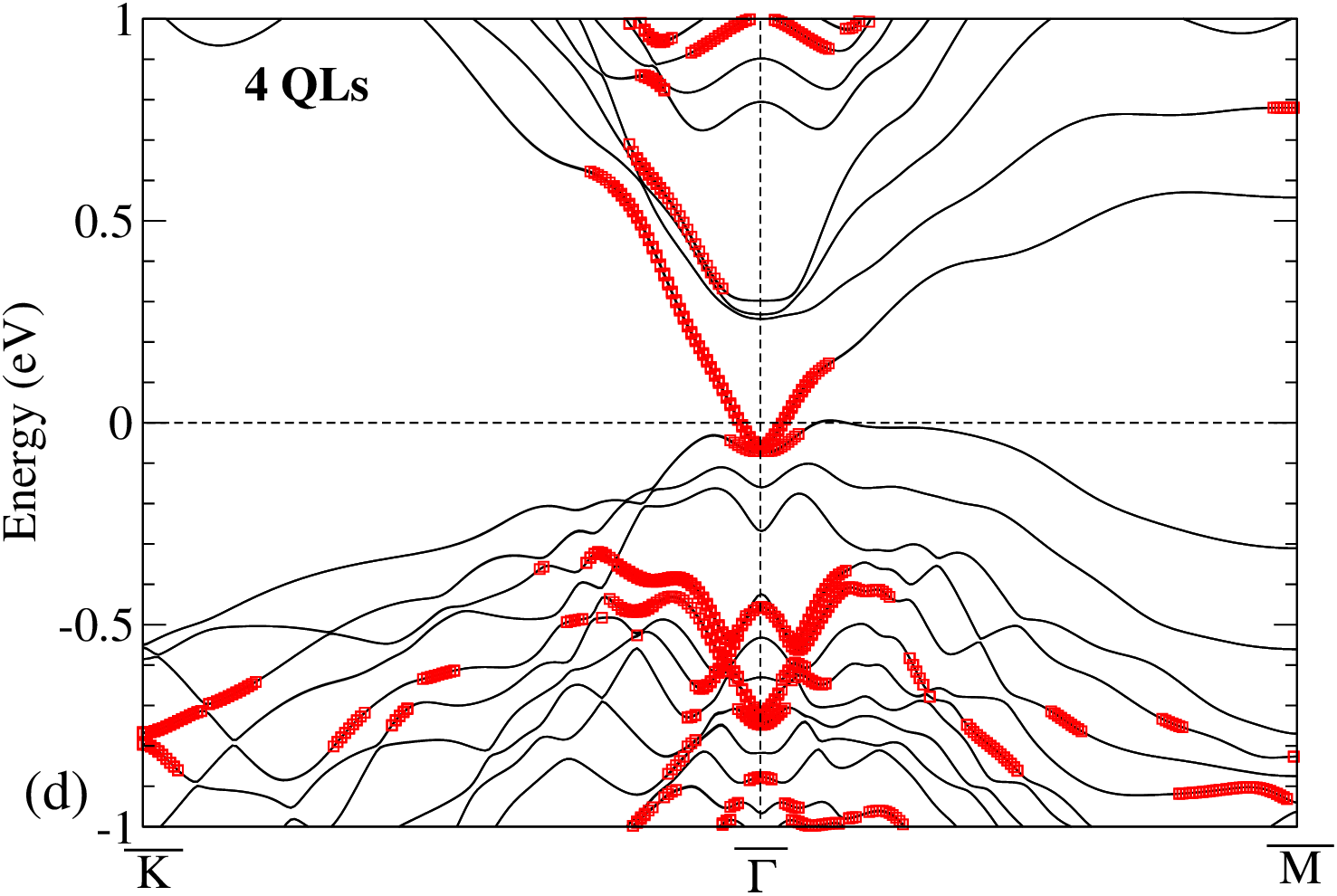}
\hspace{0.2truecm}
\includegraphics[width=5.6cm, height=4.4cm]{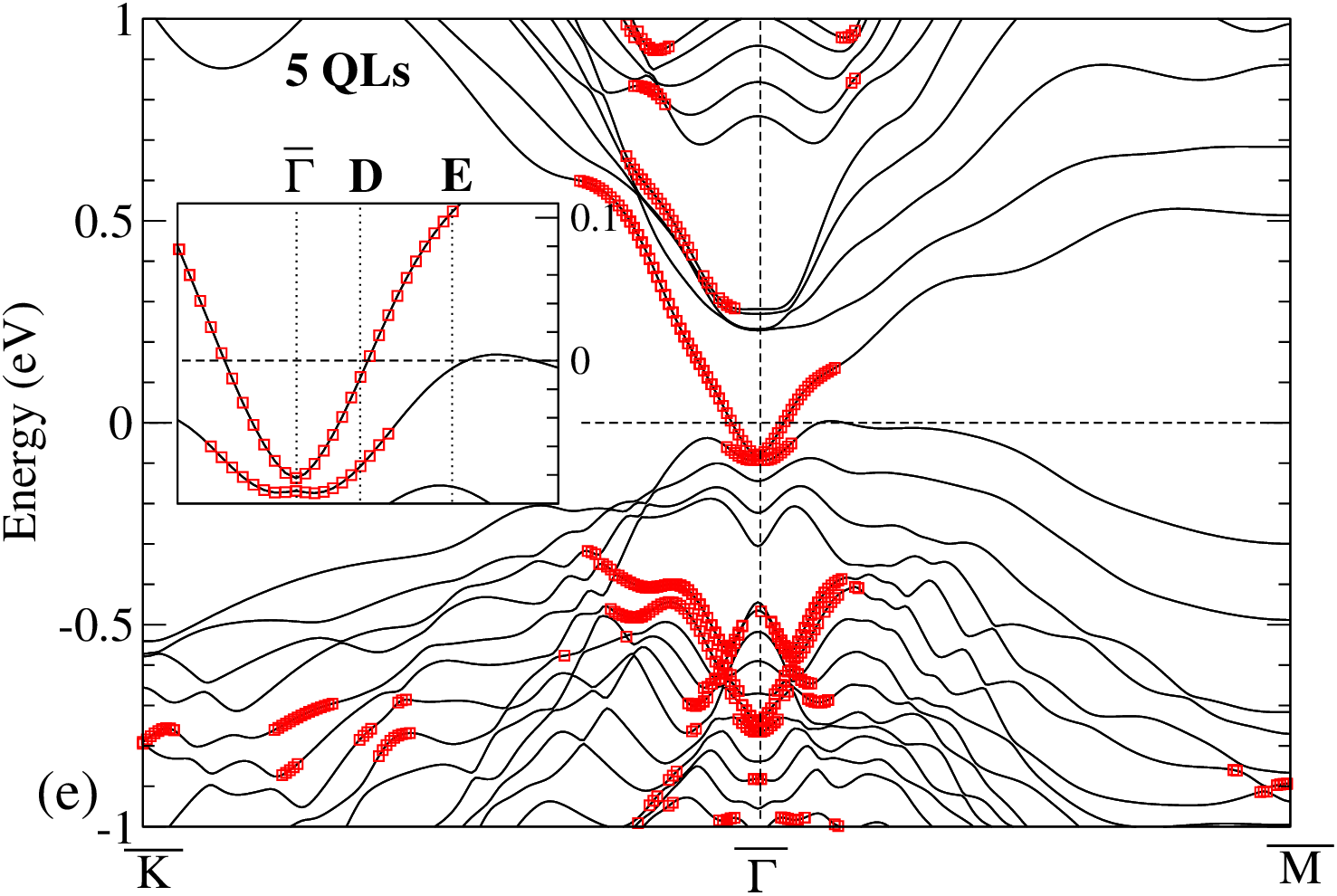}
\hspace{0.2truecm}
\includegraphics[width=5.6cm, height=4.4cm]{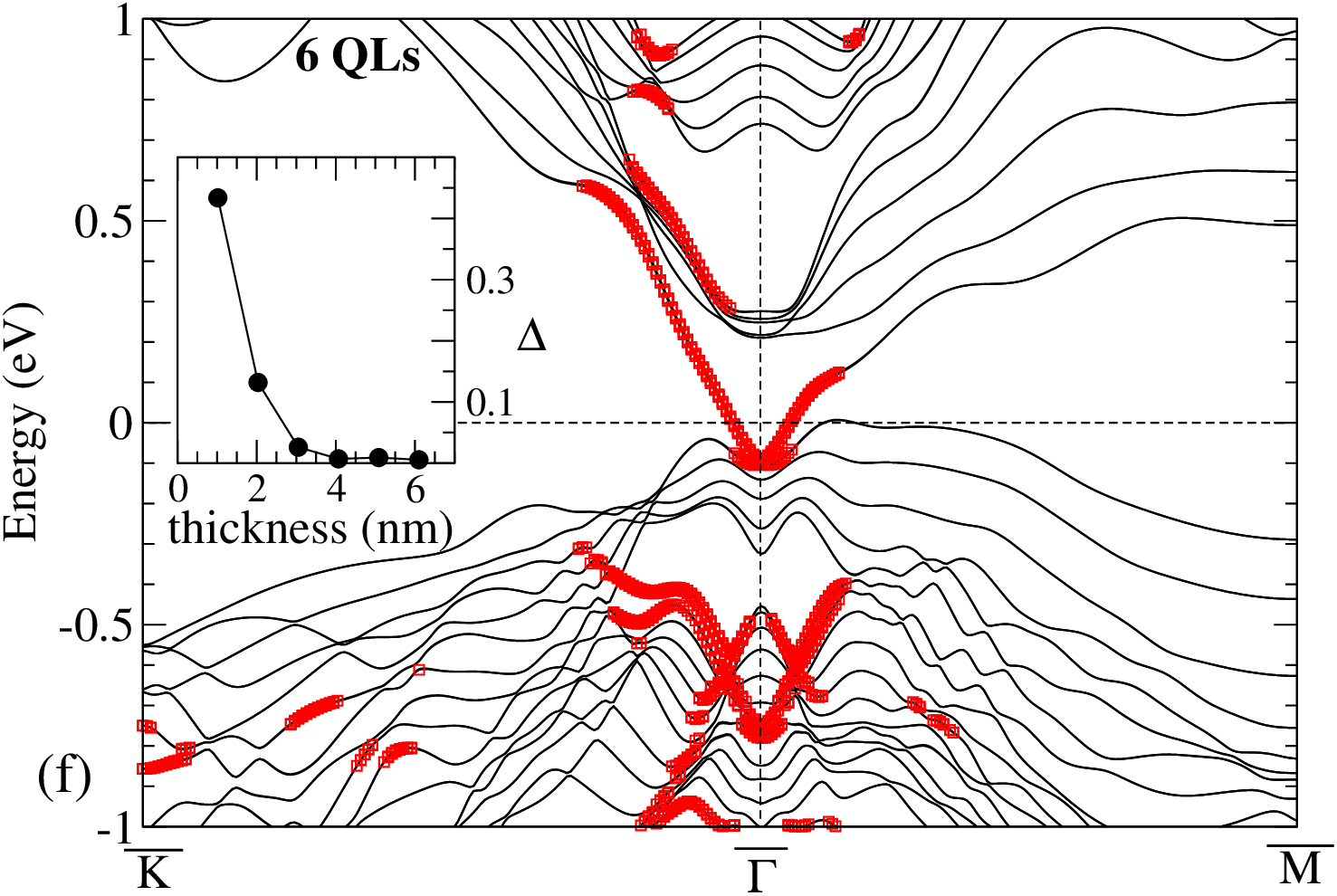}
\caption{(Color online) Band structures of Bi$_2$Te$_3$(111) films with six thicknesses
(surface states marked by the squares).
Inset in (a): Two-dimensional Brillouin zone. Inset in (b): Schematic view of one QL
of a Bi$_2$Te$_3$(111) film. A, B, and C indicate the planar sites of the Te and Bi atoms.
$z_1$=1.737~\AA,~$z_2$=2.033~\AA,~$z_3$=2.625~\AA~\cite{NAKA63}.
Inset in (c): Zoom-in of the 3 QL band structure near $E_f=0$. Inset in (e):
Zoom-in of the 5 QL band structure near $E_f=0$. Inset in (f): $\Delta$ (eV) vs thickness
(nm).}
\label{fig:slab-band-crit2}
\end{figure*}

Let us discuss the structure of a Bi$_2$Te$_3$(111) film and parameter values for band-structure
calculations of the films. Along the (111) direction (trigonal axis), a Bi$_2$Te$_3$ slab is built in units
of a QL, which consists of two Bi and three Te layers that alternate [inset in Fig.~\ref{fig:slab-band-crit2}(b)].
In each atomic layer, the Bi or Te atoms form a triangular lattice, and their in-plane positions coincide with
those in a (111) surface of a faced-centered-cubic lattice. For each QL [inset in Fig.~\ref{fig:slab-band-crit2}(b)],
the Te(1) and Bi(1)/Bi(2) layers bond more strongly than the Bi(1)/Bi(2) and Te(2) layers, since $z_1 < z_2$.
Neighboring QLs are separated by $z_3$ ($> z_2$), which causes the Te(1) layers between neighboring
QLs to interact via weak van der Waals forces, allowing exfoliation \cite{TEWE10}. For the slab electronic structure,
we use {\tt VASP} \cite{VASP} with the PBE GGA and PAW pseudopotentials as in the bulk case. We do not perform
reconstruction of the surfaces. Compared to the bulk case, the electronic structure of the slabs, especially
of surface bands, converges more slowly with parameter values ($E_c$, $N_k$, and the number of vacuum layers). 
For the six slabs with different thicknesses considered,
a large vacuum layer equivalent to 5 QLs ($=$50.829~\AA) is added. The values of $E_c$ and $N_k$ used for the six slabs
are listed in Table~\ref{table:1}. 

Self-consistent DFT calculations with SOC are first carried out using the parameter values
listed in Table~\ref{table:1} until the total energy converges to within $1 \times 10^{-6}$ eV. The two-dimensional
band structure is then computed non-self-consistently using a charge density distribution obtained from the previous self-consistent
calculations. Accuracy in the structure of the surface bands depends on the accuracy in the charge density distribution. 
Thus, extremely well-converged self-consistent calculations are required for an accurate identification of the surface bands.





We now introduce our method to identify surface states from the slab band structure.
Surface bands refer to bands localized on the top or bottom surface layers of a slab \cite{FURT96}, but
not all bands in a slab belong to the surface bands. To identify surface bands, a wave function at a given
energy band and $\vec{k}_{\parallel}$ (momentum parallel to the surface) is projected onto spherical harmonics
which are non-zero within some radius around each ion. Then two criteria are applied to the wave function
projections. In criterion 1, the surface bands are identified based on a critical percentage of the
projections onto the top two or the bottom two atomic layers. The critical percentage depends on
the slab thickness. In criterion 2, the surface bands are identified according to a critical percentage of the
projections onto the top or the bottom QL. Criterion 2 is suitable for multiple QLs. The
density of states (DOS) projected onto each atomic layer of the topmost and bottommost QL reveals
that the DOS projected onto the inner Te(1) layers more closely resembles that projected 
onto the outer Te(1) layers than those
onto the middle Te(2) layers [inset in Fig.~\ref{fig:slab-band-crit2}(b)]. Additionally, the electron density
spreads over all five atomic layers within the topmost or bottommost QL.

Band structures of Bi$_2$Te$_3$(111) slabs of 1-6 QLs, are shown in
Fig.~\ref{fig:slab-band-crit2} with the surface bands marked. Here the surface bands are identified
by applying criterion 1 to the 1 QL and 2 QL slabs and criterion 2 to the other slabs.
Critical percentages are self-consistently determined such that the wave function projections onto the 
surfaces are stable. The specific critical percentages are as follows: 45\% and 30\% for 1 QL and 2 QLs, 
respectively, and 60\%, 50\%, 40\%, and 40\% for 3 QLs, 4 QLs, 5 QLs, and 6 QLs, respectively. For very thin
slabs (1-2 QLs), criterion 2 is not applied. When criterion 1 is applied to the slabs thicker
than 2 QLs, the critical percentages are 25\%, 20\%, 16\%, and 16\% for 3 QLs, 4 QLs, 5 QLs, and 6 QLs,
respectively. Among the surface states marked in Fig.~\ref{fig:slab-band-crit2}, only those within
0.2~eV above and 0.3~eV below $E_f$ near $\overline{\Gamma}$ survive as surface states.
The surface bands outside this energy window, far away from $\overline{\Gamma}$, are strongly coupled to the
bulk bands [Fig.~\ref{fig:bulk-band}(b)]. A comparison of the surface bands identified using the criteria 
1 and 2 reveals that the surface states within the energy window remain essentially unchanged using the
different criteria. 



\begin{table}
\begin{center}
\caption{Wave function projections onto each QL (from the topmost to the bottommost QL)
for the 5 QL slab, as a function of energy band and $k_{\parallel}$.
The specific $k_{\parallel}$ points for the V-shaped (U-shaped) band shown in the inset of
Fig.~\ref{fig:slab-band-crit2}(e) are labeled as $\overline{\Gamma}_V$, D$_V$, E$_V$
($\overline{\Gamma}_U$, D$_U$, E$_U$). Each band has a double degeneracy.
Two wave functions corresponding to $\overline{\Gamma}_V$
have the following projections onto each QL with the opposite spin moments:
(0.503,0.204,0.024,0.085,0.189), (0.189,0.085,0.024,0.204,0.503).
The surface states (SS) are identified based on the criterion 2.}
\label{table:2}
\begin{ruledtabular}
\begin{tabular}{c|c|c|c|c|c|c}
  &  $\overline{\Gamma}_U$ & $\overline{\Gamma}_V$ & D$_U$ & D$_V$ &  E$_U$ & E$_V$  \\ \hline
$k_{\parallel}$ (2$\pi/a$) &  0 & 0 & 0.022 & 0.022 & 0.057 & 0.057 \\
Energy(eV)  & $-$0.091 & $-$0.082 & $-$0.074 & $-$0.012 &  $-$0.001 & 0.111  \\ \hline
topmost     & 0.438 & 0.503 & 0.490 & 0.795 &  0.328 & 0.573 \\
top$-$1    & 0.190 & 0.204 &  0.356 & 0.182 & 0.323 & 0.274  \\
middle    &  0.060 & 0.024 & 0.128 & 0.016  & 0.187 & 0.101  \\
bottom$+$1  & 0.094 & 0.085 &  0.023 & 0.000  & 0.083 & 0.038  \\
bottommost     & 0.217 & 0.189 & 0.003 & 0.000 &  0.080 & 0.017  \\ \hline
SS?     &  Yes &  Yes  & Yes  & Yes   & No    & Yes
\end{tabular}
\end{ruledtabular}
\end{center}
\end{table}

Now we present a physical interpretation of the surface states identified above. Let us first consider a very
thick slab where the top surface bands do not interact with the bottom surface bands. When the thick slab
has symmetric surfaces (the top and bottom surfaces in an identical environment), the surface bands become
fourfold degenerate at $\overline{\Gamma}$, because of Kramers degeneracy at the time-reversal
invariant momentum for both the top and bottom surfaces. Away from $\overline{\Gamma}$, the fourfold
degeneracy is lifted to a twofold degeneracy due to SOC. Neglecting the fact that
the spin-up and spin-down states are not eigenstates in the presence of SOC, for a symmetric thick slab, 
the top surface bands with spin-up (spin-down) are degenerate with the bottom surface bands with spin-down 
(spin-up).

As the slab thickness decreases, the top surface bands interact with the bottom surface bands, resulting in
an opening of a band gap $\Delta$, even at $\overline{\Gamma}$. At $\overline{\Gamma}$, a linear even combination
of the top spin-up and the bottom spin-up surface states are degenerate with a linear even combination of the top spin-down
and the bottom spin-down surface states. At $\overline{\Gamma}$, the energies of these even combinations are separated 
from those of the similar odd combinations (with a double degeneracy) by $\Delta$. This gap increases with increasing
$k_{\parallel}$. As the thickness decreases, $\Delta$ increases. Even though
the symmetric slab still possesses time-reversal and inversion symmetry, a strong interaction between the top and bottom
surfaces mixes spin-up with spin-down states, resulting in a reduction in spin polarization at each surface.
Thus, spin-flip scattering at each surface will no longer be suppressed. However, the thin slab must show
a remnant of a bulk TI: the number of $E_f$ crossings of the surface bands for a given 
surface (either the top or bottom surface).

Our calculated values of $\Delta$ at $\overline{\Gamma}$ and of the indirect band gap
$\Delta E_{\rm ind}$ corroborate robust surface states for three or more QLs.
The 1 QL and 2 QL slabs show $\Delta$ of 0.4338 and 0.1319 eV
at $\overline{\Gamma}$, respectively. This gap decreases exponentially with increasing thickness
and it saturates at a thickness of 4 QLs [Table~\ref{table:1}, inset in Fig.~\ref{fig:slab-band-crit2}(f)], in
agreement with Refs.~[\onlinecite{LI09,LIU10}].
For the slabs four or less QLs thick, $\Delta$ does not originate entirely from the surface states.
At $\overline{\Gamma}$, for the slabs three or less QLs thick, the valence band does not have a surface character 
[Fig.~\ref{fig:slab-band-crit2}(a),(b), the V-shaped band in the inset of (c)],
while for the 4 QL slab, the band slightly below the V-shaped band loses its surface character. 
However, as $k_{\parallel}$ increases, for the 3 QL (4 QL) slab, the V-shaped band (the band slightly below the V-shaped band) 
retrieves its surface character. The 1 QL and 2 QL slabs show $\Delta E_{\rm ind}$ of 0.301 and 0.057~eV, respectively. 
As the thickness increases, $\Delta E_{\rm ind}$ vanishes.

We now use our accurate identification of surface states in the slabs to examine their topological properties.
For a three-dimensional TI, $\nu_0$ is determined from the bulk band structure by a product of parity eigenvalues of
all occupied bands (counting Kramers degenerate pairs only once) at the eight time-reversal invariant
momenta \cite{FU07}. However, this procedure is not well
defined in slabs. Thus, instead, we use an equivalent criterion \cite{FU07}. The surface bands of a TI
cross $E_f$ an odd number of times between time-reversal invariant momenta ($\nu_0=1$). With the notations
in Ref.[\onlinecite{FU07}], $\nu_0$ can be obtained from
$(-1)^{\nu_0} = \pi_{\overline{\Gamma}} \pi_{\overline{M}} \pi_{\overline{M}_2} \pi_{\overline{M}_3}$,
where $\overline{\Gamma}$, $\overline{M}$, $\overline{M}_2$, and $\overline{M}_3$ are four time-reversal
invariant momenta in the two-dimensional Brillouin zone [inset in Fig.~\ref{fig:slab-band-crit2}(a)].
If the surface bands cross $E_f$ an odd (even) number of times between $\overline{\Gamma}$ and
$\overline{M}$, then $\pi_{\overline{\Gamma}} \pi_{\overline{M}} = -1$ ($+1$). Another
equivalent criterion states that the conduction and valence bands with opposite parities are inverted by SOC
at the Dirac point ($\overline{\Gamma}$) for slabs with topologically nontrivial surface states.
These two criteria must provide the same result.

For the 1 QL and 2 QL slabs, the surface states neither cross $E_f$ between $\overline{\Gamma}$
and $\overline{M}$ [Fig.~\ref{fig:slab-band-crit2}(a),(b)] nor between $\overline{M_2}$ and $\overline{M_3}$.
Thus, the 1 QL and 2 QL slabs have topologically trivial ($\nu_0=0$) surface states. This is also corroborated by
our finding that the conduction and valence bands are not inverted at $\overline{\Gamma}$ by SOC. Note that
the topologically trivial nature of the surface states would not change by 
shifting $E_f$ upward. 

Let us discuss the slabs three or more QLs thick. Their surface states
are marked in Fig.~\ref{fig:slab-band-crit2}(c)-(f). Within the small energy window from $E_f$, only two surface
bands are identified. The features of these two surface bands discussed below do 
not depend on the thickness. To better examine the surface states, for example,
for the 5 QL slab, the projections of the wave function onto each QL at three $k_{\parallel}$-points for
the two energy bands [shown in the inset of Fig.~\ref{fig:slab-band-crit2}(e)] 
are listed in Table~\ref{table:2}. The V-shaped valence band near $\overline{\Gamma}$ is a surface band
[Fig.~\ref{fig:slab-band-crit2}(c)-(f)]. This surface band intersects
$E_f$ only once between $\overline{\Gamma}$ and $\overline{M}$ [insets in Fig.~\ref{fig:slab-band-crit2}(c),(e)]
and joins the bulk bands as $k_{\parallel}$ moves further away from $\overline{\Gamma}$. The occupied U-shaped band 
slightly below the V-shaped band near $\overline{\Gamma}$ [insets in Fig.~\ref{fig:slab-band-crit2}(c),(e)], is again 
a surface band only within a very narrow window in $k_{\parallel}$, yet it loses its surface character 
before it crosses $E_f$ [see the projections of this band at the E point, E$_U$, in Table~\ref{table:2} and 
the inset of Fig.~\ref{fig:slab-band-crit2}(e)]. Additionally, none of the surface bands cross $E_f$ between 
$\overline{M}_2$ and $\overline{M}_3$. SOC indeed inverts the conduction and valence bands. 
Thus, {\it slabs three or more QLs thick have topologically non-trivial ($\nu_0$=1) surface states,
independent of the thickness.}
Our findings clearly reveal that the topological nature of the surface states persists with the thickness,
differing from previous calculations \cite{LIU10}, yet in agreement with experiment \cite{LI09}.


We have constructed a method to accurately identify topological surface states within DFT.
Using this method, we have investigated the topological nature of the surface states in thin films of Bi$_2$Te$_3$(111)
using DFT. We have found that the topological nature of the surface states remains robust with the film
thickness and that the surface states are topologically protected for films of three or more QLs.
The method and our findings are applicable to thin films of other types of TI.

\begin{acknowledgments}
The authors thank T. Stanescu for discussions. K.P. was supported by NSF DMR-0804665,
D.M. by US DOE DE-FG05-92ER40677, J.J.H. by US DOE DE-FG02-08ER46532.
Computational support was provided by Intel 64 cluster (Abe) at the NCSA under DMR060009N
and VT ARC.
\end{acknowledgments}


\end{document}